\documentclass[10pt,conference]{IEEEtran}

\IEEEoverridecommandlockouts
\usepackage[font={footnotesize}]{caption}
\usepackage{subcaption}
\usepackage{cite}
\usepackage{balance}
\usepackage{amsmath,amssymb,amsfonts}
\usepackage{algorithmic}
\usepackage{graphicx}
\usepackage{textcomp}
\usepackage{xcolor}
\usepackage{tikz}
\usepackage{pgfplots} 
\usepackage{tikzscale}
\usepackage{pgfplots}
\usepackage{cite}
\usepackage{mathtools}
\usepackage{amsmath,amssymb,amsfonts}
\usepackage{algorithmic}
\usepackage{textcomp}
\usepackage{accents} 
\usepackage{xcolor}
\def\BibTeX{{\rm B\kern-.05em{\sc i\kern-.025em b}\kern-.08em
    T\kern-.1667em\lower.7ex\hbox{E}\kern-.125emX}}
    
\begin{document}
%\title{Multi-BS sensing for 3D RIS Localization with joint angle and delay processing}
\title{RIS Position and Orientation Estimation via Multi-Carrier Transmissions and Multiple Receivers
}
% alternative
%\title{Joint RIS Position and Orientation Estimation via Multi-Carrier Transmissions and Multiple Receivers}

\author{Reza~Ghazalian\IEEEauthorrefmark{1}, Hui~Chen\IEEEauthorrefmark{2}, George~C.~Alexandropoulos\IEEEauthorrefmark{3}\IEEEauthorrefmark{4},\\ Gonzalo Seco-Granados\IEEEauthorrefmark{5}, Henk~Wymeersch\IEEEauthorrefmark{2}, and Riku~J{\"a}ntti\IEEEauthorrefmark{1}\\
\IEEEauthorrefmark{1}Aalto University, Finland, 
\IEEEauthorrefmark{2}Chalmers University of Technology, Sweden,\\ \IEEEauthorrefmark{3}National and Kapodistrian University of Athens, Greece, \IEEEauthorrefmark{4}Technology Innovation Institute, UAE,\\ \IEEEauthorrefmark{5}Universitat Autònoma de Barcelona, Spain\\
e-mails: \{reza.ghazalian, riku.jantti\}@aalto.fi, \{hui.chen, henkw\}@chalmers.se, \\alexandg@di.uoa.gr, gonzalo.seco@uab.cat\vspace{-0.5cm}
}

\maketitle
\begin{abstract}
Reconfigurable intelligent surfaces (RISs) are considered as an enabling technology for the upcoming sixth generation of wireless systems, exhibiting significant potential for radio localization and sensing. An RIS is usually treated as an anchor point with known position and orientation when deployed to offer user localization. However, it can also be attached to a user to enable its localization in a semi-passive manner. In this paper, we consider a static user equipped with an RIS and study the RIS localization problem (i.e., joint three-dimensional position and orientation estimation), when operating in a system comprising a single-antenna transmitter and multiple synchronized single-antenna receivers with known locations. We present a multi-stage estimator using time-of-arrival and spatial frequency measurements, and derive the Cramér-Rao lower bounds for the estimated parameters to validate the estimator's performance. Our simulation results demonstrate the efficiency of the proposed RIS state estimation approach under various system operation parameters.
%as the root-mean-square error of the estimations attains the corresponding CRBs. We finally show that using spatial frequency estimates at the RIS makes localization feasible with less infrastructure, i.e., RXs, compared to the method merely using TOA. 
\end{abstract}
\begin{IEEEkeywords}
Localization, reconfigurable intelligent surfaces, orientation, multi-carrier transmission, OFDM, time of arrival. \end{IEEEkeywords}
\vspace{-2mm}
\section{Introduction}
Reconfigurable intelligent surfaces (RISs) are recently considered as a promising paradigm shift for 6G wireless systems~\cite{huang2019reconfigurable,Marco2019}. An RIS is a thin planar surface comprising multiple  low-cost metamaterials whose response when excited from impinging electromagnetic waves is dynamically programmable \cite{Tsinghua_RIS_Tutorial}. This feature renders RISs as enablers of programmable signal propagation, motivating the concept of smart wireless environments~\cite{RISE6G_COMMAG}, which can be exploited for offering coverage extension as well as localization and mapping~\cite{wymeersch2020radio}.
%controllable elements capable of changing the amplitude and or phase of an electromagnetic (EM) wave and can thus achieve a controllable wireless channel~\cite{huang2019reconfigurable}. This property makes RIS beneficial not only for communication, but also for localization and mapping in terms of improved accuracy and extended physical coverage~\cite{wymeersch2020radio}.
In most cases, RISs are deployed as anchors with known locations and orientations, and can support or enable user localization; this is called \textit{RIS-aided or RIS-enabled localization} \cite{Keykhosravi2022infeasible}. In addition, RISs can be carried by a user to enable its semi-passive localization~\cite{keykhosravi2021semi}. In this case, the RIS state is unknown and needs to be estimated; this problem is known as \textit{RIS  localization} or \textit{RIS state estimation}. %(we call this RIS localization). 

In terms of {RIS-aided and RIS-enabled localization}, a large number of studies have been conducted ranging from two-dimensional (2D)~\cite{huang2022near} to three-dimensional (3D)~\cite{elzanaty2021reconfigurable} cases, far-field (FF)~\cite{keykhosravi2022ris} to near-field (NF)~\cite{abu2021near} propagation conditions, as well as from indoor~\cite{zhang2021metalocalization} to outdoor~\cite{luo2022uav} scenarios. In~\cite{huang2022near}, the authors proposed a method to discriminate the FF from NF targeting the minimization of the user localization error in an RIS-assisted positioning setup. In~\cite{elzanaty2021reconfigurable}, the Cramér-Rao lower bounds (CRBs) on the 3D position and orientation of a user for both the NF and FF cases were derived for an RIS-assisted joint communication and localization system. Considering user mobility and the spatial-wideband effect, an efficient localization method was presented for the FF scenario in~\cite{keykhosravi2022ris}. An RIS-aided multi-user localization method for an indoor scenario was proposed in~\cite{zhang2021metalocalization}, where a particle swarm optimization was deployed for the RIS phase profile design. Joint communication and user localization with an RIS mounted on an unmanned aerial vehicle was presented in~\cite{luo2022uav}, where the vehicle's trajectory and the RIS phase shifts were jointly optimized. Collectively, all the above works showcase that RISs can boost, or even enable in certain cases, localization, offering improved estimation accuracy with reduced infrastructure cost. %(e.g., localization can be performed in a SISO system with the assist of RIS~\cite{keykhosravi2022ris}).
% However, the RIS state (i.e., location and orientation) were assumed to be known in all previous studies, which has practical constraints. 

In contrast to RIS-aided or RIS-enabled localization, the problem of \emph{RIS state estimation} is more challenging as a typical passive RIS has no local estimation capabilities, and the problem has received only limited attention. As shown in~\cite{keykhosravi2021semi,ghazalian2022bi}, the localization of an RIS can be formulated as a bi-static sensing problem. In~\cite{ghazalian2022bi}, an RIS localization method was developed for a 2D scenario. %, but cannot adapt to more practical 3D scenarios. 
By using measurements for the time of arrival (TOA), a 3D RIS localization algorithm was proposed in~\cite{keykhosravi2021semi}. However, the RIS orientation cannot be estimated from TOA measurements. %, which would in principal require more TOA measurements.    
%However,this work only considers TOA information, and the estimation of RIS orientation needs more measurements%, i.e., using more RXs to make the problem feasible.
Note that an RIS contains a planar array of metamaterials, hence, the angle of arrival (AOA) and/or the angle of departure (AOD) information can be exploited for the surface's orientation estimation. However, for an RIS without sensing capabilities and unknown position and orientation, these angles are hard to estimate. 
%\footnote{AOD cannot be estimated without knowledge of the AOA, and the AOA is not known because the RIS is passive and its position and orientation are unknown. Thus, AOA and AOD can be estimated as spatial frequency, which will be defined in Sec.~\ref{sec:signal_and_channel_model}. %Since RIS is passive, AOA and AOD cannot be extracted independently and can only be estimated as spatial frequency, which will be defined in Sec.~\ref{sec:signal_and_channel_model}.
%} at the RIS can be exploited for orientation estimation.

In this paper, we study the RIS joint 3D position and orientation estimation problem under a generic FF channel for the case of a static RIS in the vicinity of multi-carrier transmissions from a single-antenna transmitter (TX) to multiple synchronized single-antenna receivers (RXs).  
%we propose a RIS localization algorithm, which exploits TOA and spatial frequency at the RIS. We consider the 3D RIS localization problem under static, synchronous, and FF scenario, using multiple receivers (RXs). 
We present a low-complexity estimator based on the measurements at RXs, including the TOA and spatial frequencies at the RIS. To validate its efficiency, we derive the theoretical CRBs for the estimated parameters, which are shown to be achieved under certain operating conditions. We compare the proposed method with that in \cite{keykhosravi2021semi} that only uses TOA measurements to show the benefits of exploiting the spatial frequency at the RIS for its orientation estimation.

\subsubsection*{Notation}
Vectors and matrices are indicated by lowercase and uppercase bold letters, respectively. The element in the $i$th row and
$j$th column of matrix $\mathbf{A}$ is denotes by $\left[\mathbf{A}\right]_{i,j}$. The sub-index
$i:j$ determines all the elements between $i$ and $j$. The complex conjugate, Hermitian, transpose, and Moore–Penrose inverse operators are represented by $\left( .\right)^*$, $\left( .\right)^{\mathsf{H}}$, $\left( .\right)^\top$, and $\left( .\right)^\dag$, respectively. $\Vert.\Vert$ calculates the norm of vectors or Frobenius norm of matrices. By $\odot$ and $\otimes$, we indicate the element-wise and Kronecker products, respectively. $\jmath=\sqrt{-1}$ and $\mathbf{1}_K$ is a column vector comprising all ones with length $K$. The functions $\text{atan2}(y,x)$ and $\text{acos}(x)$ are the four-quadrant inverse tangent and inverse cosine functions, respectively.

\section{System Model}
\begin{figure}
    \centering
    \includegraphics[width=.9\linewidth]{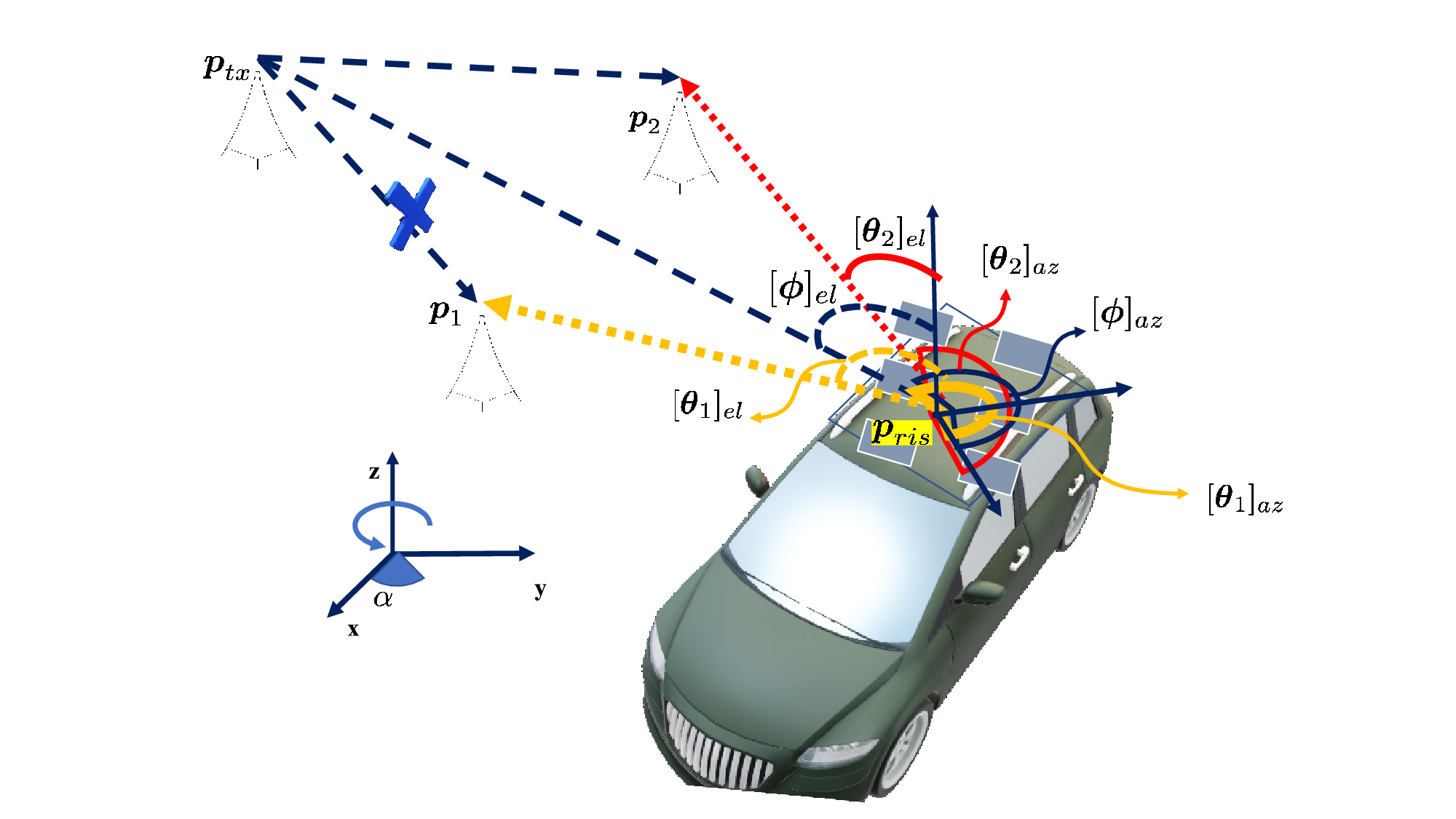}
    \caption{The considered system model comprising a single-antenna TX, $M$ single-antenna RXs (here, $M=2$ for illustrational simplicity), and an RIS with unknown 3D position and 1D orientation $\alpha$ (w.r.t.~the $z$ axis).}
    \label{fig:overview}\vspace{-5mm}
\end{figure}
In this section, we present the considered wireless system and  signal model for the proposed RIS localization problem.
% RIS-enabled 3D localization. 
% \vspace{-3mm}
\subsection{System Setup}
We consider the multi-carrier downlink transmission system illustrated in Fig.~\ref{fig:overview}, which comprises a single-antenna TX with a known location $\mathbf{p}_\text{tx}$, $M$ RXs with known locations $\mathbf{p}_1, \dots,\mathbf{p}_M$, and an RIS with an unknown state, defined as its 3D position $\mathbf{p}_\text{ris}$ together with its 1D orientation angle $\alpha$. Note that in this work, We assumed that the RIS only has one orientation degree of freedom. By considering the following rotation matrix:
\begin{equation}
\label{eq:rot-matrix}
    \mathbf{R}_{\alpha} \triangleq \begin{bmatrix} \cos{\alpha}&\sin{\alpha}& 0\\-\sin{\alpha}&\cos{\alpha}&0 \\ 0 & 0 & 1  \end{bmatrix},
\end{equation}
the direction vectors from the RIS to the TX and to each $m$th RX in the RIS local coordinate system can be obtained as follows:
% \begin{subequations}
\begin{equation}\label{eq:directions}
    \mathbf{a}_\text{tr}  = \mathbf{R}_{\alpha} \frac{\mathbf{p}_\text{tx}-\mathbf{p}_\text{ris}}{\Vert \mathbf{p}_\text{tx}-\mathbf{p}_\text{ris}\Vert }, \quad \text{and} \quad 
    % \end{equation}
    % \begin{equation}
    \mathbf{a}_\text{rm}  = \mathbf{R}_{\alpha} \frac{\mathbf{p}_m-\mathbf{p}_\text{ris}}{\Vert \mathbf{p}_m-\mathbf{p}_\text{ris}\Vert }.
\end{equation}
% \end{subequations}
% \footnote{We assume that the TX and RXs are located in the far-field range of the RIS.}
% The RIS is a uniform planar array with $K_r$ and $K_c$ elements in each row and column, respectively, with a half-wavelength inter-element space $\Delta\triangleq\lambda/2$, with $\lambda$ being the signal wavelength.
The RIS is a uniform planar array with $K_r$ and $K_c$ elements in each row and column, respectively. The inter-element space should be less than half-wavelength  $\Delta\le\lambda/2$, with $\lambda$ being the signal wavelength. By assuming that the TX and RXs are synchronized under LOS blockage conditions \footnote{Note that, when the LOS path is present, the requirement on TX-RX synchronization can be removed and the approach in~\cite{keykhosravi2021semi} can be used for separating the LOS and RIS paths, resulting in this paper's simplified LOS-blockage assumption. However, for the considered synchronized TX and RXs, the LOS path does not convey any information.}, 
%\footnote{\textcolor{red}{This is a practical assumption in factory scenario where several synchronized directional antennas are facing downwards working as TX and RXs. Besides, this is not a limiting assumption, but only invoked for notation convenience. When the LOS path is present, it can be separated from the RIS path as in~\cite{keykhosravi2021semi}. When the TX and RX are synchronized, the LOS path does not convey any information.}}
each RX receives the signals from the TX via the RIS. We assume that the RXs send their measurements to a fusion center \cite{RISE6G_COMMAG} to estimate the 3D position and 1D orientation of the RIS.   %Accordingly, the location of the element in the $\ell$-th row  $(\ell \in \{1,\dots, M_r\})$ and the $m$-th column $(m \in \{1,\dots, M_c\})$ is $\mathbf{p}_{\ell,m} = [(\ell-(M_r-1)/2)\Delta, 0 , (m-(M_c-1)/2)\Delta]^\top$ in the RIS local coordinate system. 
\vspace{-1mm}
\subsection{Signal and Channel Models}
\label{sec:signal_and_channel_model}
The TX transmits $T$ orthogonal frequency division multiplexing (OFDM) symbols over time via $N_c$ sub-carriers. We select $T$ sufficiently small such that the considered channel is constant during each transmission interval. Without loss of generality, we assume that all the transmitted symbols over all sub-carriers have unit power. %To increase the number of observations,
In addition, RIS is programmed to change its phase profile randomly during each discrete time instant $t$. The RIS phase profile is denoted by vector $\boldsymbol{\gamma}_t \in \mathbb{C}^{K\times1}$, where $|[\boldsymbol{\gamma}_t]_k| = 1$ $\forall$$k$ and $K\triangleq K_r K_c$. 

After removing the cyclic prefix and computing the fast Fourier transform (FFT), the baseband received signal $\mathbf{Y}_m \in \mathbb{C}^{N_c\times T}$ can be mathematically expressed as:\footnote{For convenience, we don't consider the LOS path, which, if it is present and resolvable, can be removed from the observation. Since TX and RXs are synchronized, the LOS path provides no additional information. }
%\vspace{-3mm}
% Regarding these assumptions, and after removing cyclic prefix and taking fast Fourier transform (FFT), one can write the received signals at the $m$th as the following matrix, i.e., $\mathbf{Y}_m \in \mathbb{C}^{N_c\times T}$:
\begin{equation}\label{eq:observation_mth-rx}
    \mathbf{Y}_m \triangleq g_m \sqrt{P_t} \mathbf{d}(\,\tau_m )\,\mathbf{b}^\top(\,\boldsymbol{\theta}_m,\boldsymbol{\phi})\, \boldsymbol{\Gamma} + \mathbf{W}_m , 
\end{equation}
where $g_m \triangleq \rho_m e^{\jmath \varphi_m}$ is an unknown channel gain, and
% , modeled with $\varphi_m \sim \mathcal{U} [0,2\pi)$ and amplitude \cite[eqs. (27) and (29)]{ellingson2021path}:
% \begin{equation}\label{eq:chan_gain}
%     \rho_m = \frac{\lambda^2 (\,\cos{[\boldsymbol{\theta}_m]_\text{el}} \cos{[\boldsymbol{\phi]}_\text{el}})\,^{0.285}}{16 \pi \Vert\mathbf{p}_\text{tx}-\mathbf{p}_\text{ris}\Vert \Vert\mathbf{p}_m-\mathbf{p}_\text{ris}\Vert}, 
% \end{equation}
$\boldsymbol{\theta}_m \triangleq [[\boldsymbol{\theta}_m]_\text{el} , [\boldsymbol{\theta}_m]_\text{az}]^\top$ and $\boldsymbol{\phi} \triangleq [[\boldsymbol{\phi}]_\text{el} , [\boldsymbol{\phi}]_\text{az}]^\top$ are the AOD from the RIS to the $m$th RX and the AOA at the RIS from the TX, respectively. In fact, $\boldsymbol{\theta}_m$ is the angle in the direction of vector $\mathbf{a}_\text{rm}$, where
\begin{align}
    [\boldsymbol{\theta}_m]_\text{el} & = \text{acos}([\mathbf{a}_\text{rm}]_3) \label{eq:elevation}\\
    [\boldsymbol{\theta}_m]_\text{az}& = \text{atan2}([\mathbf{a}_\text{rm}]_2,[\mathbf{a}_\text{rm}]_1) \label{eq:azimuth}.
\end{align}
%$[\boldsymbol{\theta}_m]_\text{el}= \text{acos}([\mathbf{a}_\text{rm}]_3/\Vert\mathbf{a}_\text{rm}\Vert)$, and $[\boldsymbol{\theta}_m]_\text{az}= \text{atan2}([\mathbf{a}_\text{rm}]_2,[\mathbf{a}_\text{rm}]_1)$. 
Similarly, $\boldsymbol{\phi}$ is the angle associated with vector $\mathbf{a}_\text{tr}$. In addition, $P_t$ denotes the transmission power and $\mathbf{W}_m \in \mathbb{C}^{N_c\times T}$ is the noise matrix
containing zero-mean circularly-symmetric independent and
identically distributed Gaussian elements with variance\footnote{We assume that the noise variances at the all RXs are the same. The extension to distinct values for the variances is straightforward.} 
%\footnote{We assume that the noise variances at the all RXs are the same. The results can be extended to the case that noise levels is different at the RXs, i.e., $\sigma^2_1 = \dots = \sigma^2_M = \sigma^2$, where $\sigma^2_m$ shows the noise variance at the $m$th RX receiver.} 
$\sigma^2$. The delay steering vector $\mathbf{d}(\cdot)$ is defined as follows:%\footnote{Note that, in our observation model \eqref{eq:observation_mth-rx}, we have ignored the beam squint effect, due to our assumption that the wavelength remains relatively constant over the transmission bandwidth.}:
\begin{equation}\label{eq:delay_tau}
    \mathbf{d}(\,\tau_m)\, \triangleq [ 1,e^{-\jmath 2\pi \Delta f \tau_m} \dots, e^{-\jmath 2\pi (\,N_c-1)\,\Delta f \tau_m}\, ] \,^\top,
\end{equation}
where $\Delta f$ is the sub-carrier spacing. The delay of the propagation path is $\tau_m \triangleq (\Vert\mathbf{p}_\text{tx}-\mathbf{p}_\text{ris}\Vert+ \Vert\mathbf{p}_m-\mathbf{p}_\text{ris}\Vert)/c$ with $c$ being the speed of light. In \eqref{eq:observation_mth-rx}, $\boldsymbol{\Gamma} \triangleq [\boldsymbol{\gamma}(0),\dots,\boldsymbol{\gamma}(T)]^\top$ and $\mathbf{b}(\cdot) \in  \mathbb{C}^{K\times1}$ is the Hadamard product of the RIS array responses from both propagation sides, which is defined as:
\begin{equation}\label{eq:b_vect}
\mathbf{b} (\,\boldsymbol{\theta}_m,\boldsymbol{\phi})\, \triangleq  \boldsymbol{a}(\,\boldsymbol{\theta}_m)\odot\boldsymbol{a}(\,\boldsymbol{\phi}\,)\,,
\end{equation}
%\vspace{-4mm}
where for an arbitrary elevation and azimuth pair $\boldsymbol{\psi}$, we define $\boldsymbol{a}(\boldsymbol{\psi}) \triangleq \boldsymbol{a}_r(\boldsymbol{\psi})\otimes \boldsymbol{a}_c(\boldsymbol{\psi})$. The $n$th element of vectors $\mathbf{a}_r$ and $\mathbf{a}_c$ are defined as follows:
\begin{subequations} \label{eq:RISreponses}
\begin{equation}\label{eq:ar}
 [\boldsymbol{a}_r (\boldsymbol{\psi})]_n\triangleq e^{-\jmath \frac{2\pi n\Delta}{\lambda} \sin{[\boldsymbol{\psi}]_\text{el}}\cos{[\boldsymbol{\psi}]_\text{az}}} , n=0,\ldots, K_r-1
\end{equation}
\begin{equation}\label{eq:ac}
 [\boldsymbol{a}_c (\boldsymbol{\psi})]_n\triangleq e^{-\jmath \frac{2\pi n\Delta}{\lambda} \sin{[\boldsymbol{\psi}]_\text{el}}\sin{[\boldsymbol{\psi}]_\text{az}}}, n=0,\ldots, K_c-1.
 \end{equation}
\end{subequations}
Considering the latter definition of $\boldsymbol{a}(\boldsymbol{\psi})$ and given \eqref{eq:b_vect}--\eqref{eq:RISreponses}, the following expression is deduced:
\begin{equation}\label{eq:b_modify}
\mathbf{b} (\,\boldsymbol{\theta}_m,\boldsymbol{\phi})\, =  \mathbf{a}_0 (\,\boldsymbol{\theta}_m,\boldsymbol{\phi}) \otimes \mathbf{a}_1 (\,\boldsymbol{\theta}_m,\boldsymbol{\phi})\, ,
\end{equation}
where the $n$th element of $\mathbf{a}_0(\cdot)$ and $\mathbf{a}_1(\cdot)$ can be expressed as:
\begin{subequations} \label{eq:a0_a1}
\begin{equation}\label{eq:a0}
 [\,\mathbf{a}_0 (\,\boldsymbol{\theta}_m,\boldsymbol{\phi})\,]\,_n= e^{-\jmath \frac{2\pi n\Delta}{\lambda}\omega_0(\,\boldsymbol{\theta}_m,\boldsymbol{\phi})\,}  ,
\end{equation}
\begin{equation}\label{eq:a1}
  [\,\mathbf{a}_1 (\,\boldsymbol{\theta}_m,\boldsymbol{\phi})\,]\,_n= e^{-\jmath \frac{2\pi n\Delta}{\lambda}\omega_1(\,\boldsymbol{\theta}_m,\boldsymbol{\phi})\,},
 \end{equation}
\end{subequations}
where we have defined the spatial frequencies %$\omega_0(\cdot)$ and $\omega_1(\cdot)$:
\begin{subequations} \label{eq:spatial_freq}
\begin{equation}\label{eq:w0}
 \omega_0(\,\boldsymbol{\theta}_m,\boldsymbol{\phi})\,\triangleq \sin{\boldsymbol{[\phi}]_\text{el}}\cos{[\boldsymbol{\phi}]_\text{az}}+\sin{\boldsymbol{[\theta}_m]_\text{el}}\cos{[\boldsymbol{\theta}_m]_\text{az}}
\end{equation}
\begin{equation}\label{eq:w1}
 \omega_1(\,\boldsymbol{\theta}_m,\boldsymbol{\phi})\,\triangleq \sin{\boldsymbol{[\phi}]_\text{el}}\sin{[\boldsymbol{\phi}]_\text{az}} +\sin{\boldsymbol{[\theta}_m]_\text{el}}\sin{[\boldsymbol{\theta}_m]_\text{az}}  .
 \end{equation}
\end{subequations}
To ease notation, we hereafter define $\omega_0^{m}\triangleq\omega_0(\,\boldsymbol{\theta}_m,\boldsymbol{\phi})$ and $\omega_1^{m}\triangleq\omega_1(\,\boldsymbol{\theta}_m,\boldsymbol{\phi})$. Accordingly,  $\mathbf{b}(\boldsymbol{\theta}_m,\boldsymbol{\phi})$ can be written as a function of $\boldsymbol{\omega}^m = [\omega_0^{m},\omega_1^{m}]^\top$, i.e., as $\mathbf{b}(\boldsymbol{\omega}^m)$.

Based on the observations \eqref{eq:observation_mth-rx}, we next present an analysis for the considered parameter estimation problem (i.e., the RIS's 3D position and 1D orientation) together with a low-complexity estimation scheme.
%This localization task aims at estimating the 3D position and 1D orientation of the RIS based on received signals at the RXs with the known TX and RXs positions and RIS phase profile.

\section{Proposed Estimation Methodology}
In this section, we describe how to estimate the RIS position and orientation from \eqref{eq:observation_mth-rx}, starting with a Fisher information analysis. 
\subsection{Fisher Information Analysis}
We derive the Fisher information matrix (FIM) and the CRB for the unknown channel (i.e., $\tau_m$, $\omega_0^{m}$, $\omega_0^{m}$, $\rho_m$, and $\phi_m$) and RIS-state parameters (i.e., $\mathbf{p}_\text{ris}$ and $\alpha$).
% RIS location and orientation estimation. We first derive corresponding bounds on the channel
% parameters, namely, $\tau_m$, $\omega_0^{m}$, $\omega_0^{m}$, $\rho_m$ and $\phi_m$. Then,
% we transform these bounds into the RIS state parameters, i.e., $\mathbf{p}_\text{ris}$ and $\alpha$.
To this end, we introduce the noise-free part of the observation stacked at the fusion center, say $\boldsymbol{M} \in \mathbb{C}^{N_c M\times T}$ given \eqref{eq:observation_mth-rx}, \eqref{eq:b_modify}, \eqref{eq:w0}, and \eqref{eq:w1} as $\boldsymbol{M} \triangleq [\boldsymbol{M}_\text{1}^\top,\dots, \boldsymbol{M}_M^\top]^\top$, where $\boldsymbol{M}_m\triangleq g_m \sqrt{P_t} \mathbf{d}(\,\tau_m )\,\mathbf{b}^\top (\,\boldsymbol{\omega}^{m})\, \boldsymbol{\Gamma} \quad \forall m$.
% \begin{equation}\label{eq:mu}
%     \boldsymbol{\mu} \triangleq \begin{bmatrix}  g_1 \sqrt{P_t} \mathbf{d}(\,\tau_1 )\,\mathbf{b}^\top (\,\boldsymbol{\omega}^{1})\, \boldsymbol{\Gamma}\\
%     \vdots \\ 
%     g_M \sqrt{P_t} \mathbf{d}(\,\tau_M )\,\mathbf{b}^\top (\,\boldsymbol{\omega}^{M})\, \boldsymbol{\Gamma}
% \end{bmatrix}
% \end{equation}
We also introduce the $5M\times1$ vector with the unknown channel parameters $\boldsymbol{\eta_{\text{ch}}}\triangleq [\boldsymbol{\eta}^\top, \boldsymbol{\rho}^\top,\boldsymbol{\varphi}^\top]^\top$, $\boldsymbol{\eta}\triangleq [\boldsymbol{\tau}^\top, \boldsymbol{\omega}_0^\top, \boldsymbol{\omega}_1^\top]^\top\in \mathbb{R}^{3M\times1}$, and the $4\times 1$ vector with the RIS state $\boldsymbol{\zeta} \triangleq [\mathbf{p}_{\text{ris}}^\top, \alpha]^\top$, where $\boldsymbol{\tau}\triangleq [\tau_1, \dots,\tau_M]^\top$ , $\boldsymbol{\omega}_0\triangleq [\omega_0^1,\dots,\omega_0^M]^\top$, $\boldsymbol{\omega}_1\triangleq [\omega_1^1,\dots,\omega_1^M]^\top$, $\boldsymbol{\rho}\triangleq[\rho_0,\dots,\rho_M]^\top$, and $\boldsymbol{\varphi}\triangleq[\varphi_0,\dots,\varphi_M]^\top$. Based on \cite[Sec. 3.9]{kay1993fundamentals}, the FIM $\mathbf{J}_{\boldsymbol{\eta}_{\text{ch}}}\in \mathbb{R}^{\text{5M}\times\text{5M}}$ is defined as follows:
% \begin{equation}\label{eq:FIM1}
% % \mathbf{J}(\,\boldsymbol{\mu})\,= \frac{2}{\sigma^2} \[\sum_{t=1}^T\]\[\sum_{n_c=1}^N_c 1\]
% \mathbf{J}(\boldsymbol{\eta})= \frac{2}{\sigma^2} \sum_{t=1}^T\sum_{n_c=1}^{N_c} \Re \Bigg\{\frac{ \partial [\boldsymbol{\mu}]_{t,n_c}}{\partial\boldsymbol{\eta}} \left( \frac{ \partial [\boldsymbol{\mu}]_{t,n_c}}{\partial\boldsymbol{\eta}}\, \right)^{\mathsf{H}}\, \Bigg\}.
%  \end{equation}
 \begin{equation}\label{eq:FIM1}
% \mathbf{J}(\,\boldsymbol{\mu})\,= \frac{2}{\sigma^2} \[\sum_{t=1}^T\]\[\sum_{n_c=1}^N_c 1\]
\mathbf{J}_{\boldsymbol{\eta}_{\text{ch}}}\triangleq \frac{2}{\sigma^2} \sum_{t=1}^T \Re \Big\{\Big(\frac{ \partial [\boldsymbol{M}]_{:,t}}{\partial\boldsymbol{\eta}_{\text{ch}}} \Big) ^{\mathsf{H}}  \frac{ \partial [\boldsymbol{M}]_{:,t}}{\partial\boldsymbol{\eta}_{\text{ch}}}\,\Big\}.
 \end{equation}
We can then respectively derive the CRB related to $\tau_m$, $\omega_0^m$, and $\omega_1^m$ at $m$th RX, namely, the time error bound  ($\text{TEB}_m$), the first spatial frequency bound ($\text{WEB}_0^m$), and the second spatial frequency bound ($\text{WEB}_1^m$) as:
 %\begin{subequations}\label{eq:chan_EB}
 \begin{equation}\label{TEB}
      \sqrt{\mathbb{E}[\left( \tau_m -\hat{\tau}_m \right)^2  ]}\geq \text{TEB}_m\triangleq  \sqrt{[ \mathbf{J}_{\boldsymbol{\eta}_{\text{ch}}}^{-1}]_{m,m}} \quad,
 \end{equation}
 %\end{subequations}
 and for $i \in \{1,2\},$
\begin{align}\label{W0EB}
  \sqrt{\mathbb{E} [\left( \omega_{i-1}^m -\hat{\omega}_{i-1}^m \right)^2  ]}&\geq \text{WEB}_{i-1}^m\triangleq  \sqrt{[ \mathbf{J}_{\boldsymbol{\eta}_{\text{ch}}}^{-1}]_{m+iM,m+iM}}%\nonumber
\end{align}
%   \begin{equation}\label{W0EB}
%       \sqrt{\mathbb{E}\left [\left( \omega_0^m -\hat{\omega}_0^m \right)^2  \right]}\geq \text{WEB}_0^m\triangleq  \sqrt{\left[ \mathbf{J}(\boldsymbol{\eta}_{\text{ch}})^{-1}\right]_{m+M,m+M}} \quad,
%  \end{equation}
%  \begin{equation}\label{W1EB}
%       \sqrt{\mathbb{E}\left [\left( \omega_1^m -\hat{\omega}_1^m \right)^2  \right]}\geq \text{WEB}_1^m\triangleq  \sqrt{\left[ \mathbf{J}(\boldsymbol{\eta}_{\text{ch}})^{-1}\right]_{m+2M,m+2M}} \quad,
%  \end{equation}
where $m\in\{1,\dots,M\}$, and $\hat{\tau}_m$, $\hat{\omega}_0^m$, and $\hat{\omega}_1^m$ are the estimations of the true parameters $\tau_m$, $\omega_0^m$, and $\omega_1^m$, respectively. To derive the FIM of the RIS state, we apply a transformation from the channel parameter vector $\boldsymbol{\eta}$ to the variables in the state vector $\boldsymbol{\zeta}$. Then, we obtain the equivalent FIM (EFIM) of
$\boldsymbol{\eta}$ as $\mathbf{J}_{\boldsymbol{\eta}}= [[\mathbf{J}_{\boldsymbol{\eta}_{\text{ch}}}^{-1}]_{1:3M,1:3M}]^{-1}$. Accordingly, the FIM of RIS states is defined by means of the transformation matrix $\mathbf{T}\in \mathbb{R}^{3M\times 4}$ (i.e., the Jacobian) as $\mathbf{J}_{\boldsymbol{\zeta}} = \mathbf{T}^\top \mathbf{J}_{\boldsymbol{\eta}}\mathbf{T}$. The $\ell$th row and $q$th entry of $\mathbf{T}$ is defined as \cite[eq. (3.30)]{kay1993fundamentals}:
 \begin{equation}\label{eq:Jacobian_Mat}
      [\mathbf{T}]_{\ell,q} \triangleq \frac{\partial [\,\boldsymbol{\eta}]\,_\ell}{\partial [\,\boldsymbol{\zeta}]\,_q}.
 \end{equation}
The derivations of $\mathbf{J}_{\boldsymbol{\eta}}$ and $\mathbf{T}$ are provided in the Appendix.

Capitalizing on \eqref{eq:FIM1} and \eqref{eq:Jacobian_Mat}, the desired position error bound (PEB) and orientation error bound (OEB) are computed:
 \begin{subequations}
\begin{align}\label{eq:PEB}
 \sqrt{\mathbb{E} [ \Vert\mathbf{p}_{\text{ris}} -\hat{\mathbf{p}}_{\text{ris}}\Vert^2   ]}& \geq \text{PEB}\triangleq  \sqrt{\mathrm{tr}([ \mathbf{J}_{\boldsymbol{\zeta}}^{-1}]_{1:3,1:3})},
\\\label{eq:OEB} \sqrt{\mathbb{E} [ (\alpha -\hat{\alpha})^2   ]}& \geq \text{OEB}\triangleq  \sqrt{[ \mathbf{J}_{\boldsymbol{\zeta}}^{-1}]_{4,4}}, \end{align}
\end{subequations}
where $\hat{\mathbf{p}}_{\text{ris}}$ and $\hat{\alpha}$ are the estimates of the true RIS position and orientation, respectively.
\vspace{-1mm}
\subsection{Maximum Likelihood Estimator}
We stack all observations at each RX in the matrix $\mathbf{Y}$, i.e., $\mathbf{Y} \triangleq [{\mathbf{Y}_1}^\top, \dots , {\mathbf{Y}_M}^\top]^\top$ and define $\mathbf{g}\triangleq [{g}_1, \dots , {g}_M]^\top$. By using the definition of $\boldsymbol{M}$ and \eqref{eq:observation_mth-rx}--\eqref{eq:w1}, we can define the maximum likelihood estimator (MLE) as follows: 
\begin{align}\label{eq:MLE}
  \left[\hat{\mathbf{g}},\hat{\alpha},\hat{\mathbf{p}}_{\text{ris}} \right] &\triangleq  \arg\max_{\mathbf{g},\alpha, \mathbf{p}_{\text{ris}}} f\left( \mathbf{Y} |\mathbf{g} , \alpha, \mathbf{p}_{\text{ris}}  \right)\\ \nonumber
  &= \arg\min_{\mathbf{g}, \alpha, \mathbf{p}_{\text{ris}}} \Vert \mathbf{Y} - \boldsymbol{M} (\,\mathbf{g},\alpha, \mathbf{p}_{\text{ris}})\,\Vert^2.
\end{align}
To solve \eqref{eq:MLE}, we can adopt gradient descent methods (e.g., Newton's method). However, this is challenging since the objective function is non-convex having many local optima. In addition, the gradient descent methods are sensitive to the initial point, i.e., they can get trapped in local optima without appropriate initial points. To this end, we next present a low-complexity estimator to find a proper initial guess. 

\subsection{Low-Complexity Estimator}
%In this section, we develop a multi-stage estimator with low complexity for finding an initial estimate of the RIS location and orientation. 
The estimator first estimates the TOA and spatial frequencies at each RIS. Then these estimates are combined to determine the RIS state.  
\subsubsection{TOA Estimation  at each RX} Let $\mathbf{F} \in \mathbb{C}^{N_F \times N_c}$ be the inverse FFT matrix, defined as $[\mathbf{F}]_{\ell,q} = (1/N_F) e^{\jmath 2\pi \ell q/N_F}$
%$\forall \ell,q=1,2,\ldots,N_c$
. By computing $\mathbf{Z}_m = \mathbf{F}\mathbf{Y}_m$ at each RX, we can coarsely estimate the TOA as 
%For TOA estimation at each RX, we take IFFT of the column of the matrix $\mathbf{Y}_m$, i.e., $\mathbf{Z}_m = \mathbf{F}\mathbf{Y}_m$, where $\mathbf{F} \in \mathbb{C}^{N_F \times N_c}$ is the IFFT matrix. The $\ell$th row and $q$th element of the IFFT matrix is $\mathbf{F}_{\ell,q} = (1/N_F) e^{\jmath 2\pi \ell q/N_F}$ , where $N_F$ denotes the length of IFFT  vector. Given $\mathbf{Z}$, we then can estimate TOA coarsely as
 %\begin{equation}\label{eq:TOA_estimate}
     $\tilde{k}_m \triangleq \arg \max_{k} \Vert \mathbf{f}^\top_k \mathbf{Z}_m\Vert,$
 %\end{equation}
%where $\mathbf{f}_k$ is vector comprising $N_F$ zeros, except with one in the $k$th entry. 
where $\mathbf{f}_k$ is a vector comprising all zeros and a one at its $k$th entry. Next, by solving $\tilde{\delta}_m\triangleq \arg \max_{\delta_m \in \left[0 , 1/\left(N_F \Delta f\right)\right]} \Vert \mathbf{f}_k^\top \mathbf{F} \left( \mathbf{Y}_m\odot \mathbf{d}\left(\delta_m \right)\mathbf{1}_T^\top\right)\Vert$ via the quasi-Newton method using ${\delta_m} =0$ as the initial point, a refined TOA estimate $\tau_m$ can be obtained \cite{keykhosravi2021semi}:
\begin{equation}
\label{eq:tau}
  \hat{\tau}_m = \frac{\tilde{k}}{N_{F}\Delta f}-\tilde{\delta}_m.
\end{equation} 
% \begin{align}
%     \tilde{\delta}= \arg \max_{\delta \in \left[0 , 1/\left(N_F \Delta f\right)\right]} \Vert \mathbf{g}^\top \mathbf{Z} \odot \left(\mathbf{d}\left(\delta \right)\mathbf{1}_T^\top\right)\Vert,
% \end{align}

  \begin{figure*}[t!]
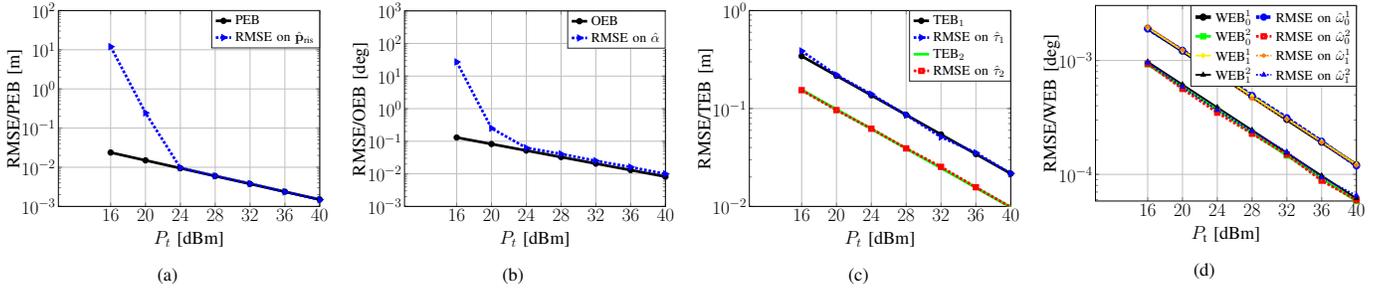

\begin{subfigure}[h!]{0.24\textwidth}
    \centering
    \includegraphics[width=\linewidth]{results/PEB_Pt.tikz}
    \caption{}
    \label{fig: POS_SNR}
\end{subfigure}
 \hfill
\begin{subfigure}[h!]{0.24\textwidth}
    \centering
    \includegraphics[width=\linewidth]{results/OEB_Pt.tikz}
    \caption{}
    \label{fig: OEB_SNR}
\end{subfigure}
 \hfill
\begin{subfigure}[h!]{0.24\textwidth}
    \centering
    \includegraphics[width=\linewidth]{results/TEB1_Pt.tikz}
    \caption{}
    \label{fig: tau1_SNR}
\end{subfigure}
 \hfill
\begin{subfigure}[h!]{0.24\textwidth}
    \centering
    \includegraphics[width=\linewidth]{results/WEB1R1_Pt.tikz}
    \caption{}
    \label{fig: tau2_SNR}
\end{subfigure}
% \begin{subfigure}[h!]{0.24\textwidth}
%     \centering
%     \includegraphics[width=\linewidth]{results/w1rx1_power.tikz}
%     \caption{}
%     \label{fig: w1r1}
% \end{subfigure}
%  \hfill
% \begin{subfigure}[h!]{0.24\textwidth}
%     \centering
%     \includegraphics[width=\linewidth]{results/w1rx2_power.tikz}
%     \caption{}
%     \label{fig: w1r2}
% \end{subfigure}
%  \hfill
% \begin{subfigure}[h!]{0.24\textwidth}
%     \centering
%     \includegraphics[width=\linewidth]{results/w2rx1_power.tikz}
%     \caption{}
%     \label{fig: w2r1}
% \end{subfigure}
%  \hfill
% \begin{subfigure}[h!]{0.24\textwidth}
%     \centering
%     \includegraphics[width=\linewidth]{results/w2rx2_power.tikz}
%     \caption{}
%     \label{fig: w2r2}
% \end{subfigure}
\caption{The evaluation of the RMSE of the estimated channel parameters and RIS state versus the transmit power $P_t$: %The location of the first RX and second RX are $[0\text{m},5\text{m},0\text{m}]^\top$, $[0\text{m},-5\text{m},0\text{m}]^\top$, respectively.
%for the simulation parameters given in Table~\ref{table: tab1}: 
(a) RMSE of the RIS position and PEB; (b) RMSE of the RIS orientation (deg) and OEB; (c) RMSE of the TOA and $\text{TEB}$; and (d) RMSE of the $\boldsymbol{\omega}^m$ and $\text{WEB}$.}\vspace{-5mm}
\label{fig:RMSE_PT}
\end{figure*}
\begin{figure}
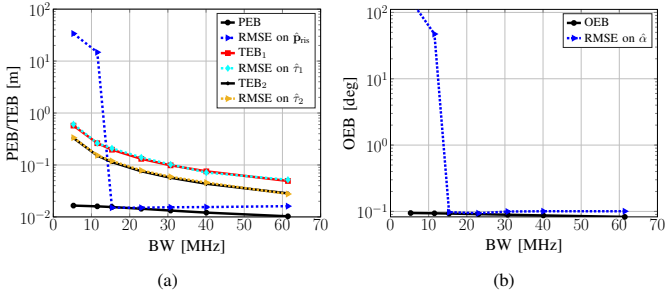
%[h!]
\begin{subfigure}%[h!]
{0.24\textwidth}
    \centering
    \includegraphics[width=\linewidth]{results/PEB_BW.tikz}
    \caption{}
    \label{fig: PEB_BW}
\end{subfigure}
\begin{subfigure}%[h!]
{0.24\textwidth}
    \centering
    \includegraphics[width=\linewidth]{results/OEB_BW.tikz}
    \caption{}
    \label{fig:OEb_BW}
\end{subfigure}
\caption{The effect of the bandwidth (BW) on the estimation accuracy for $P_\text{t} = 20 \text{dBm}$: % and the parameters given in Table~\ref{table: tab1}: 
(a) RMSE of the RIS position/TOA and PEB; and (b) RMSE of the RIS orientation (deg) and OEB.}
\label{fig:CRB_BW}
\end{figure}
% \begin{figure}
%     \centering
%     \includegraphics[width=.6\linewidth]{results/RIS_size_PEB_OEB.tikz}
%     \caption{The effect of the RIS size on the RIS state estimation accuracy for $P_\text{t} = 28 \text{dBm}$.}\vspace{-5mm}
%     \label{fig:RIS_size_effect}
% \end{figure}
\begin{figure}
    \centering
    \includegraphics[width=.6\linewidth]{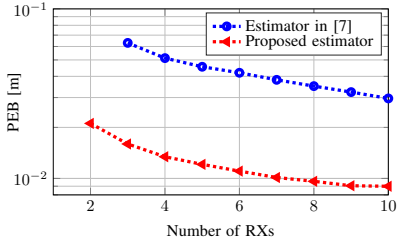}
    \caption{The comparison of the PEB using only TOA and using TOA along with AOA/AOD information to/from the RIS versus the number of RXs. %The simulation setup is given in \ref{table: tab1}.
    The RXs are equally distributed on a circle centered at the TX position with a radius equal to $5\text{m}$; this circle lies on the plane $z=0$.}\vspace{-5mm}
    \label{fig:study_w-o_aod}
\end{figure}
 
 \subsubsection{Estimating $\boldsymbol{\omega}^m$ at each RX} One can remove the effect of $\tau_m$ from $\mathbf{Y}_m$ to obtain the following expression: 
 \begin{align}\label{eq:remove_tau}
  \mathbf{Y}_m^r &= \mathbf{Y}_m \odot \left(\mathbf{d} (\,-\hat{\tau}_m)\, \mathbf{1}_T^\top\right)\\
  \nonumber
  & \approx g_m \sqrt{P_t} \mathbf{1}_{N_c} \mathbf{b}^\top(\,\boldsymbol{\omega}^{m})\, \boldsymbol{\Gamma} + {\mathbf{W}_m^d} ,
\end{align}
where $\mathbf{W}_m^d\triangleq {\mathbf{W}_m} \odot \left(\mathbf{d}\left( -\hat{\tau}_m\right) \mathbf{1}_T^\top\right) $. Next, summing $\mathbf{Y}_m^r$ over its rows, and then, computing the transpose, yields:
\begin{align}\label{eq:sum-subcar}
\mathbf{y}_m^r & = {\mathbf{Y}_m^r}^\top \mathbf{1}_{N_c}  = N_c g_m \sqrt{P_t} \boldsymbol{\Gamma}^\top \mathbf{b}(\,\boldsymbol{\omega}^{m})\, + \mathbf{w}_m^t .
\end{align}
Finally, $\boldsymbol{\omega}^m$ can be obtained from the MLE estimator:
\begin{align}\label{eq:MLE-omega}
\hat{\boldsymbol{{\omega}}}^m &= \arg\min_{\boldsymbol{\omega}^m} \Vert \mathbf{y}_m^r - N_c  \sqrt{P_t} \hat{g}_m (\,\boldsymbol{\omega}^m)\, \mathbf{\Gamma}^\top \mathbf{b} (\,\boldsymbol{\omega}^m)\,\Vert^2,
\end{align}
where $\hat{g}_m (\,\boldsymbol{\omega}^m)\, \triangleq  \mathbf{y}_m^r(\mathbf{\Gamma}^\top \mathbf{b} (\,\boldsymbol{\omega}^m))^\dag /(N_c\sqrt{P_t})$. 
% \begin{align}\label{eq:g_esimate}
%     \hat{g}_m (\,\boldsymbol{\omega}^m)\,  &= \frac{\mathbf{b}^{\mathsf{H}} (\,\boldsymbol{\omega}^m)\,\boldsymbol{\Gamma}^* \mathbf{y}_m^r}{N_c  \sqrt{P_t}\Vert \mathbf{b}^{\mathsf{H}} (\,\boldsymbol{\omega}^m)\,\boldsymbol{\Gamma}^* \Vert^2} \quad.
% \end{align}
This problem can be solved via a  2D search over the interval $[-2,2]$, providing a coarse estimation for $\boldsymbol{\omega}^m$. The estimation can be then refined via the quasi-Newton method using the latter coarse estimate as the initial point. 
% Note, in the refinement, coarse estimate is an initial point for quasi-Newton method.
\subsubsection{RIS 3D Position and 1D Orientation Estimation} Using the estimation $\hat{\tau}_m$, the RIS is constrained to lie on the intersection of $M$ spheroids defined as $\Vert \mathbf{p}_{\text{ris}} - \mathbf{p}_{\text{tx}}\Vert + \Vert\mathbf{p}_{\text{ris}} - \mathbf{p}_m \Vert = c\hat{\tau}_m$. %In this study, we consider a practical and simple scenario where TX is assumed to be located at origin and two RXs are positioned at $y$-axis, i.e., $[0, y_m, 0 ]^\top$ \footnote{This scenario can happen in the machine type communication in a factory, e.g., quality control robot tracking.}.Based on this assumption, the intersection of these two spheroids is a circle. The equation of the circle is (please see the Appendix):
% \begin{equation}\label{eq:intersection_cicle}
%     x^2 + z^2 = r^2  \quad,
% \end{equation}
% Considering \eqref{eq:intersection_cicle} and \eqref{eq:eq_sol} (See the Appendix), the candidate RIS location can be easily parameterized as  $\mathbf{p}_\text{ris} (\nu) = [r\cos{\nu} \quad y \quad r\sin{\nu}]^\top$, where $\nu \in [0, \pi]$ \footnote{This search space for $\nu$ is reasonable since we assumed that the RIS location is on top of the robot, while the locations of TX and RXs are on the ceil.}. Note that $r$ is a function the $\tau_m$, the constant $y$ plane \footnote{The intersection of two spheroids (i.e., the circle) lies on this plane.}, the locations of RXs and TX. 
In the general case, the intersection of $M>2$ spheroids can be found by the method from \cite[eq.~(16)--(18)]{keykhosravi2021semi}. However, for $M=2$, this method does not apply, as there is a one-dimensional manifold of solutions. To characterize these,  we first mesh the surface of one of the spheroids and then calculate the distances between each of the points on the mesh with the other spheroid, using the technique given in  \cite[Sec 3]{eberlyElipsoidPoint}. We finally select the candidate RIS points as the intersections, which have a distance less than a threshed $d_\text{th}$, which is manually set. The resulting set of candidate positions is denoted by $\mathcal{P}$ (where for $M>2$, this set contains a single point).

By considering that \eqref{eq:directions} depends on the RIS position and orientation, and substituting the definitions of elevation and azimuth angles 
\eqref{eq:elevation}, \eqref{eq:azimuth} in 
\eqref{eq:spatial_freq}, we notice that for a given 
$\tilde{\mathbf{p}}_{\text{ris}}\in \mathcal{P}$, $\boldsymbol{\omega}^m$ is solely  a function of $\alpha$. Hence,  $\alpha(\tilde{\mathbf{p}}_{\text{ris}})$ can be found by a line search
\begin{align}
    \hat{\alpha}(\tilde{\mathbf{p}}_{\text{ris}}) = \arg \min_{\alpha \in [0,2\pi)} \sum_{m=1}^M\Vert \hat{\boldsymbol{{\omega}}}^m-{\boldsymbol{{\omega}}}^m(\alpha)\Vert^2.
\end{align}
For $M=2$, we finally find a unique estimate of the position  %$\accentset{\ast}{\mathbf{p}} \triangleq[\hat{\mathbf{p}}_{\text{ris}} , \quad\hat{\alpha}]^\top$
via solving the following optimization problem:
% \eqref{eq:intersection_cicle} and \textcolor{blue}{ \eqref{eq:nonlin_alpha}, }the optimum solution of $\nu$, which determines the optimum RIS location and orientation, can be obtained through solving the following optimization problem:
% \begin{equation}
%     \hat{\nu} = \Vert \sum_{m = 1}^2 \mathbf{y}_m^r - \Vert^2,
% \end{equation}
% \begin{equation}\label{eq:nu_optim}
%     \hat{\nu} = \arg\min_{\nu}\Vert  \mathbf{y}^r - \boldsymbol{\mu}_0 (\,\nu)\, \Vert^2,
% \end{equation}
\begin{equation}\label{eq:nu_optim}
    \hat{\mathbf{p}}_{\text{ris}} = \arg\min_{\tilde{\mathbf{p}}_\text{ris} \in \mathcal{P}}\Vert  \mathbf{y}^r - \boldsymbol{m}_0 (\,\tilde{\mathbf{p}}_\text{ris}, \hat{\alpha}(\tilde{\mathbf{p}}_{\text{ris}}))\, \Vert^2,
\end{equation}
where $\mathbf{y}^r\triangleq\sum_{m = 1}^M \mathbf{y}_m^r$,  $\boldsymbol{m}_0({\mathbf{p}}_\text{ris},\alpha) \triangleq\sum_{m = 1}^M N_c \hat{g}_m(\,{\mathbf{p}_\text{ris}}, {\alpha})\, \sqrt{P_t} \boldsymbol{\Gamma}^\top \mathbf{b}(\,{\mathbf{p}_\text{ris}}, {\alpha})\,$, $\hat{g}_m(\,{\mathbf{p}_\text{ris}}, {\alpha})\, =  \mathbf{y}_m^r(\mathbf{\Gamma}^\top \mathbf{b} ({\mathbf{p}_\text{ris}}, {\alpha}))^\dag /(N_c\sqrt{P_t})$. %, and $\digamma$ shows the RIS candidate state set. 
%The problem \eqref{eq:nu_optim} can be solved through a search over set $\digamma$. 
Note that this state estimation can serve as 
%Finally, the estimation provided by \eqref{eq:nu_optim} refines as 
an initial guess the optimization in~\eqref{eq:MLE}. 
\begin{table}[b]
\caption{Simulation Parameters.}
\vspace{-2mm}
\begin{center}
\begin{tabular}{l c c} 
 \hline \hline
 Parameter & Symbol & Value  \\  
 \hline\hline
 Wavelength & $\lambda$ & $1 ~\text{cm}$\\
 RIS element spacing & $\Delta$ & $0.25 ~\text{cm}$ \\Light speed & $c$ & $3\times 10^{8}~ \text{m}/\text{sec}$\\
 Number of sub-carriers & $N_c$& $128$\\
 Number of transmissions & $T$& $100$\\
 Sub-carrier spacing & $\Delta f$& $120~\text{kHz}$\\
 Noise PSD&$N_0$& $-174~\text{dBm/Hz}$\\ RX's noise figure& $n_f$ & $5~\text{dB}$\\ IFFT Size & $N_F$& $4096$\\ The first RX position& $\mathbf{p}_1$& $\left[-3\text{m},5\text{m},-1\text{m}\right]^\top$\\ The second RX position& $\mathbf{p}_2$& $\left[3\text{m},-3\text{m},0\text{m}\right]^\top$ \\TX position& $\mathbf{p}_{\text{tx}}$& $\left[0\text{m},0\text{m},0\text{m}\right]^\top$ \\RIS position& $\mathbf{p}_{\text{ris}}$& $\left[4\text{m},1\text{m},-4\text{m}\right]^\top$\\
 RIS orientation& $\alpha$& $\pi/6~\text{rad}$\\
 \hline\hline
\end{tabular}
\label{table: tab1}
\end{center}
\end{table}
% \vspace{-3mm}
 \section{Numerical Results}
% \begin{itemize}
% \item Kamran work and mine
% \item RMSE/CRB vs SNR
% \item BW study
% \item RIS size study
% \item contour plot
% \item profile optimization
% \end{itemize}
%\subsection{Scenario and Parameters}
In this section, we evaluate the proposed estimator for $M=2$ receivers, and compare it with the corresponding bound. In particular, we compare the root mean square error (RMSE) of the estimated parameters with the derived CRBs, using $500$ noise realizations. The number of RIS elements was set to $K=17\times 17$, and the phase profile of each of them was drawn from the uniform distribution $[0,2\pi)$. The channel gain $g_m \triangleq \rho_m e^{\jmath \varphi_m}$ modeled with $\varphi_m \sim \mathcal{U} [0,2\pi)$ and amplitude \cite[eqs. (27) and (29)]{ellingson2021path}:
\begin{equation}\label{eq:chan_gain}
    \rho_m = \frac{\lambda^2 (\,\cos{[\boldsymbol{\theta}_m]_\text{el}} \cos{[\boldsymbol{\phi]}_\text{el}})\,^{0.285}}{16 \pi \Vert\mathbf{p}_\text{tx}-\mathbf{p}_\text{ris}\Vert \Vert\mathbf{p}_m-\mathbf{p}_\text{ris}\Vert}. 
\end{equation}
The rest of the simulation parameters are given in Table \ref{table: tab1}.
% \footnote{Note that we have set the simulation parameters such that the RXs and TX are located in the FF range of RIS.}.
% \subsubsection{RIS Position and Orientation Estimation}

\subsection{Results and Discussion}
\subsubsection{Impact of Transmission Power and Bandwidth}
In Fig.~\ref{fig:RMSE_PT}, we study the effect of the transmitted power $P_t$ on the performance of the proposed estimator. As shown, the RMSE of channel parameters ($\hat{\tau}_1$, $\hat{\tau}_2$, $\hat{\omega}_0^m$, and $\hat{\omega}_1^m$) and the RIS state ($\hat{\mathbf{p}}_{\text{ris}}$ and $\hat{\alpha}$) are decreasing functions of $P_t$, and %. The RMSE of the proposed estimator 
the CRB of the RIS state estimation is attained when $P_t\ge 24$ dBm. %As can be seen, the RIS state estimation accuracy is very sensitive to the TOA estimations. 
It can be also seen that the bottleneck for the RIS state estimation is the TOA estimation at lower transmit power, where a small TOA error leads to a large positioning error. %, which depends on the available bandwidth (BW). 
To verify this behavior, we study the effect of the signal bandwidth (BW) in Fig~\ref{fig:CRB_BW}. It is observed that, for low BW values, the localization algorithm fails, whereas more BW yields more accurate TOA estimation. 

% We also investigate the effect of the RIS size on the RIS state estimation, shown in~Fig.\ref{fig:RIS_size_effect}. As can be seen, as the number of the RIS elements increases, the error bound on the RIS state estimation decreases. That is because increasing the number of reflecting elements leads to an increase not only in the receiving signals but spatial frequency estimation at the RIS.
 
 \subsubsection{Gain of Spatial Frequency Estimation over \cite{keykhosravi2021semi}}
In Fig.~\ref{fig:study_w-o_aod}, we compare with the scenario presented in \cite{keykhosravi2021semi}, which only estimates the RIS 3D location using TOA. As can be seen, using spatial frequency estimation, which contains AOA/AOD information, along with TOA at RIS, results in a more accurate RIS localization than that using only the TOA. Besides, the estimator proposed in \cite{keykhosravi2021semi} cannot estimate the RIS location with only $M=2$ RXs, while the proposed method renders the problem indentifiable and provides an accurate solution.  
\begin{figure}[!t]
\begin{subfigure}[h!]{0.24\textwidth}
    \centering
    \includegraphics[width=\linewidth]{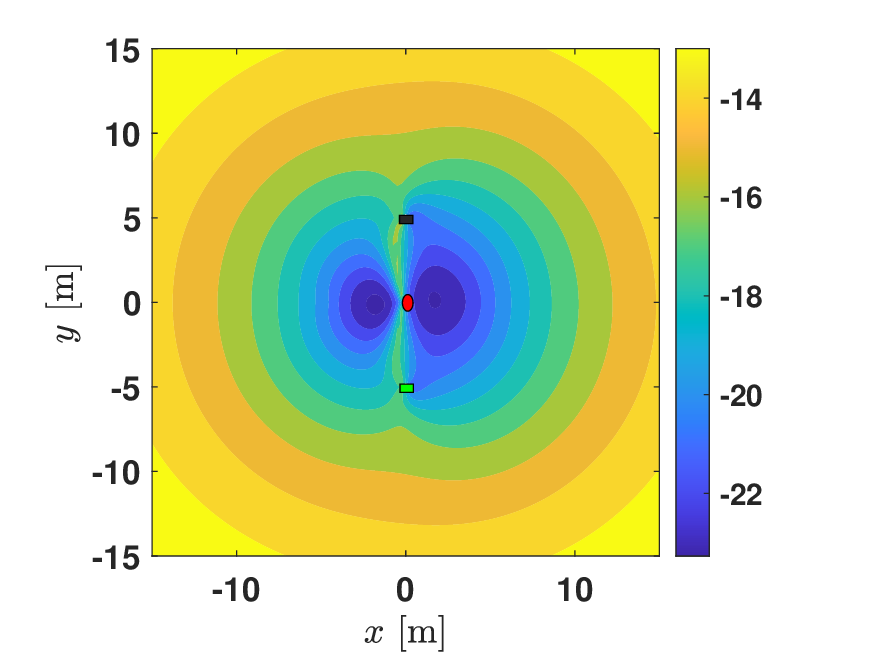}
    \caption{$10\text{log}10(\text{PEB})$, $\alpha = 0^\circ$}
    \label{fig: cont_peb0}
\end{subfigure}
 \hfill
\begin{subfigure}[h!]{0.24\textwidth}
    \centering
    \includegraphics[width=\linewidth]{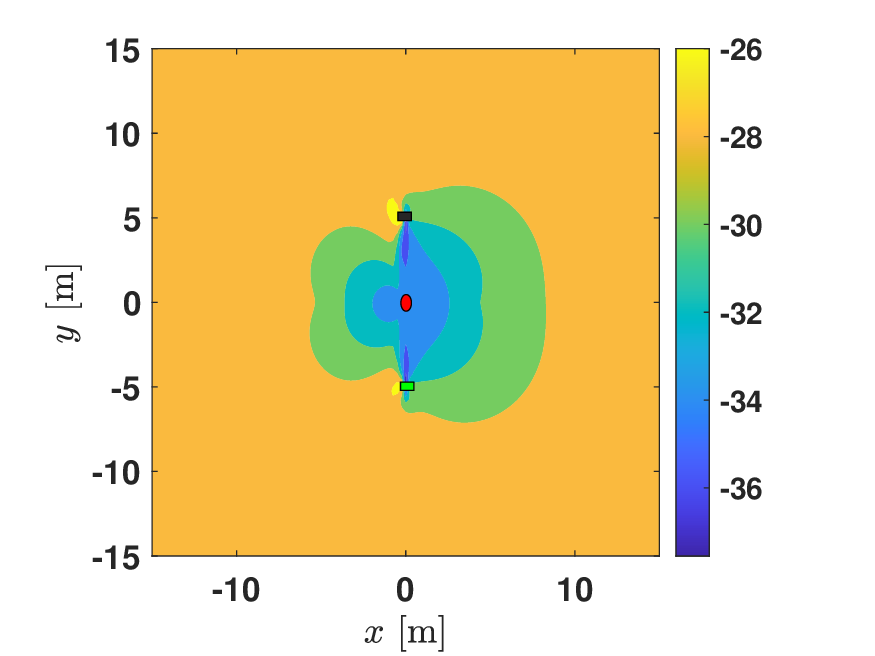}
    \caption{ $10\text{log}10(\text{OEB})$, $\alpha = 0^\circ$}
    \label{fig: cont_OEB0}
\end{subfigure}
 \hfill
\begin{subfigure}[h!]{0.24\textwidth}
    \centering
    \includegraphics[width=\linewidth]{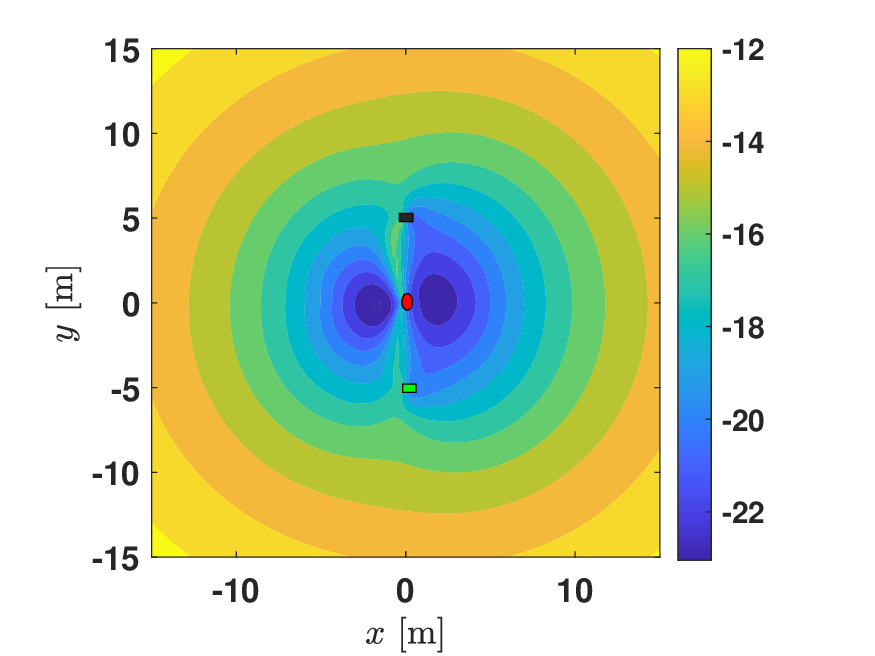}
    \caption{$10\text{log}10(\text{PEB})$, $\alpha = 30^\circ$}
    \label{fig: cont_peb30}
\end{subfigure}
 \hfill
\begin{subfigure}[h!]{0.24\textwidth}
    \centering
    \includegraphics[width=\linewidth]{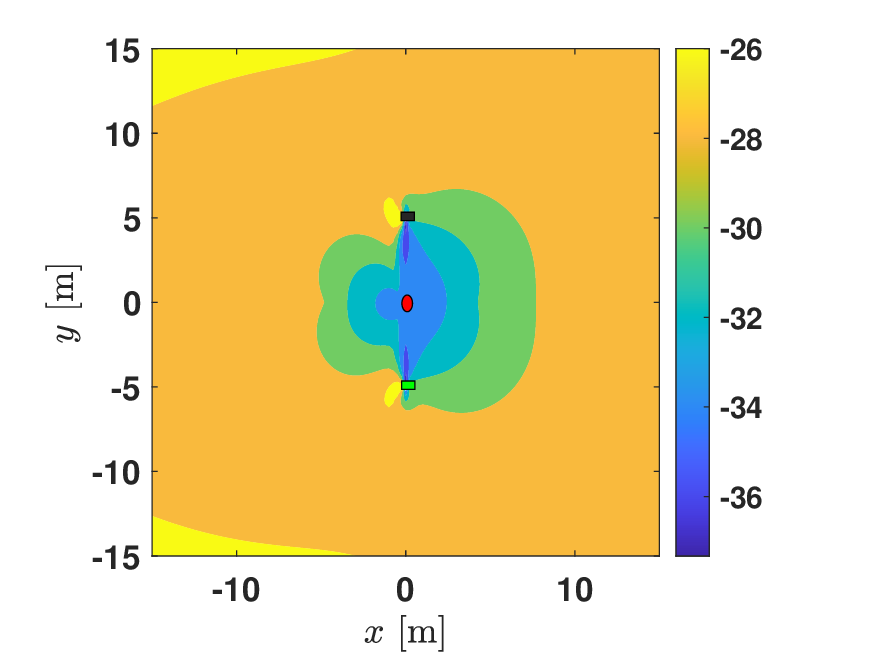}
    \caption{$10\text{log}10(\text{OEB})$, $\alpha = 30^\circ$}
    \label{fig: cont_OEB30}
\end{subfigure}
\caption{The contour plots of the PEB and OEB versus the RIS location for one
 random RIS phase profile and $P_\text{t} = 34~\text{dBm}$. The TX, first RX, and second RX positions are marked by a red ellipse, green square, and black square, respectively.}\vspace{-5mm}
\label{fig:contour}
\end{figure}
\subsubsection{Coverage Analysis}
We assess the RIS localization coverage and performance through contour plots of the PEB and OEB in Fig.~\ref{fig:contour}, when the $x$ and $y$ coordinates of the RIS are varying, while its orientation and $z$ coordinate are fixed; we particularly set $z=-1\text{m}$. Two RIS orientations were considered, namely $\alpha = 0^\circ$ and $30^\circ$, and the RXs were located on the ceiling (i.e., the plane $z=0$) in $[0\text{m},-5\text{m},0\text{m}]^\top$ and $[0\text{m},5\text{m},0\text{m}]^\top$, respectively. 
%In Fig.~\ref{fig:contour}, we also evaluate the effect of the RIS orientation. To this end, we depict contour plots for two cases where RIS has different orientations including $\alpha = 0 $ (rad) and $\alpha = \pi/6 $ (rad). 
It can be observed that high accurate localization can be obtained when the RIS is close to the TX, due to the high signal-to-noise ratio (SNR). However, one can estimate the RIS orientation with high accuracy when the RIS is located on a line connecting one of the RXs with the TX.

% \begin{figure}
%     \centering
%     \includegraphics[width=0.5\linewidth]{results/PEB_BW.tikz}
%     \caption{The comparison of the PEB using only TOA and using TOA along with AOA/AOD information at RIS versus the number of RXs. The simulation setup is given in \ref{table: tab1}. The RXs are equally distributed on the circle centered at the TX position with a radius equal to $5\text{m}$. This circle lies on the plane $z=0$.}
%     \label{fig:study_w-o_aod}
% \end{figure}
% \section{Conclusion}

%\vspace{-3mm}
\section{Conclusion}
In this paper, we presented a multi-stage estimator for the 3D position and 1D orientation estimation of an RIS in a multi-carrier system with a single-antenna TX and multiple single-antenna RXs. The proposed estimator leverages the TOA and spatial frequency measurements to estimate the unknown parameters. We showed that the RMSEs of the estimations attain the corresponding CRBs within a specific SNR range. We also demonstrated that using additional measurements, specifically the spatial frequency at the RIS, not only improves the localization accuracy compared to methods using only TOAs, but makes the RIS orientation estimation feasible with even two RXs. In future work, we will investigate the robustness of the estimator to the multipath. 

%\vspace{-3mm}
\vspace{-.8mm}
\section*{Acknowledgments}
This work has been funded in part by the Academy of Finland Profi-5 under grant number 326346, the ULTRA project under grant number 328215, the EU H2020 RISE-6G project under grant number 10101701, and in part by the Spanish Ministry of Science and Innovation Projects under grant PID2020-118984GB-I00.
\vspace{-.8mm}
\section*{Appendix}
\subsection{Derivation of FIM $\mathbf{J}_{\boldsymbol{\eta}}$}
We calculate the term $\partial [\boldsymbol{M}]_{:,t}/\partial\boldsymbol{\eta}$, using \eqref{eq:FIM1}, to derive $\mathbf{J}$. We first define vector $\mathbf{e}_{(m-1)M+1:mM}$ with length $M$, whose entries from $(m-1)M+1$ to $mM$ are equal to one and the others are equal to zero. To ease notation, we express $(m-1)M+1:mM$ via $m$, i.e., $\mathbf{e}_m$. Given the definition of $\boldsymbol{M}$, we have $\partial[\boldsymbol{M}]_{:,t}/\partial\rho_m= e^{\jmath \varphi_m} \sqrt{P_t} \mathbf{d}(\,\tau_m )\,\mathbf{b} (\,\boldsymbol{\omega}^m)\, \boldsymbol{\gamma}_t\mathbf{e}_m$, $\partial[\boldsymbol{M}]_{:,t}/\partial\varphi_m= \jmath g_m \sqrt{P_t} \mathbf{d}(\,\tau_m )\,\mathbf{b} (\,\boldsymbol{\omega}^m)\, \boldsymbol{\gamma}_t\mathbf{e}_m$, and 
\begin{equation}\label{eq:partial_tau}
    \frac{\partial[\boldsymbol{M}]_{:,t}}{\partial\tau_m}=g_m\sqrt{P_t} \frac{\partial\mathbf{d}(\,\tau_m )\,}{\partial\tau_m}\mathbf{b} (\,\boldsymbol{\omega}^m)\, \boldsymbol{\gamma}_t \mathbf{e}_m \quad, 
\end{equation}
where $ \partial\mathbf{d}(\,\tau_m )\,/\partial\tau_m = -\jmath2\pi\Delta f g_m \, (\,\mathbf{L}_c \odot
    \,\mathbf{d}(\,\tau_m)\, )\,\sqrt{P_t}\mathbf{b} (\,\boldsymbol{\omega}^m)\, \boldsymbol{\gamma}_t \mathbf{e}_m$ and
% \begin{align}\label{eq:mu_partial_d_tau}
%     \frac{\partial\mathbf{d}(\,\tau_m )\,}{\partial\tau_m}&=-\jmath2\pi\Delta f g_m \, \big(\,\mathbf{L}_c \odot
%     \,\mathbf{d}(\,\tau_m)\, \big)\,\sqrt{P_t}\mathbf{b} (\,\boldsymbol{\omega}^m)\, \boldsymbol{\gamma}_t \mathbf{e}_m,
% \end{align}
$\mathbf{L}_c \triangleq\big(\,[\,0, \dots,M-1]^\top$. In addition, for $i=0$ and $1$, we have the derivatives:
\begin{align}\label{eq:mu_partial_w1}
    \frac{\partial[\mathbf{\boldsymbol{M}}]_{:,t}}{\partial\omega_i^m}&=g_m \mathbf{d}(\,\tau_m)\,\sqrt{P_t} \frac{\partial\mathbf{b}^\top (\,\boldsymbol{\omega}^m)\,}{\partial \omega_i^m} \boldsymbol{\gamma}_t \mathbf{e}_m,
\end{align}
%\begin{subequations}
%\begin{align}\label{eq:mu_partial_w0}
%    \frac{\partial[\mathbf{\boldsymbol{M}}]_{:,t}}{\partial\omega_0^m}&=g_m \mathbf{d}(\,\tau_m)\,\sqrt{P_t} ,\frac{\partial \mathbf{b}^\top (\,\boldsymbol{\omega}^m)\,}{\partial \omega_0^m} \boldsymbol{\gamma}_t \mathbf{e}_m ,
%\end{align}
%\begin{align}\label{eq:mu_partial_w1}
%    \frac{\partial[\mathbf{\boldsymbol{M}}]_{:,t}}{\partial\omega_0^m}&=g_m \mathbf{d}(\,\tau_m)\,\sqrt{P_t} \frac{\partial\mathbf{b}^\top (\,\boldsymbol{\omega}^m)\,}{\partial \omega_1^m} \boldsymbol{\gamma}_t \mathbf{e}_m,
%\end{align}
%\end{subequations}
where $\partial \mathbf{b}(\,\boldsymbol{\omega}^m)/\partial \omega_0^{m}= -\jmath2\pi\Delta/\lambda (\, \mathbf{L} \otimes \mathbf{1})\,\odot \mathbf{b} (\,\boldsymbol{\omega}^m)\,$,
$\partial \mathbf{b}(\,\boldsymbol{\omega}^m)/\partial \omega_1^{m}= -\jmath2\pi\Delta/\lambda (\,  \mathbf{1} \otimes\mathbf{L})\,\odot \mathbf{b} (\,\boldsymbol{\omega}^m)\,
$, and 
$\mathbf{L}\triangleq[\,-\frac{M-1}{2},\dots,\frac{M-1}{2} ]\,^\top$.

% \begin{subequations}
% \begin{align}\label{eq:b_partial_w0}
%   \frac{\partial \mathbf{b}}{\partial \omega_0^{m}}&= \frac{-\jmath2\pi\Delta}{\lambda} (\, \mathbf{L} \otimes \mathbf{1})\,\odot \mathbf{b} (\,\boldsymbol{\omega}^m)\,,
% \end{align}
% \begin{align}\label{eq:b_partial_w1}
%      \frac{\partial \mathbf{b}}{\partial \omega_1^{m}}&= \frac{-\jmath2\pi\Delta}{\lambda} (\,  \mathbf{1} \otimes\mathbf{L})\,\odot \mathbf{b} (\,\boldsymbol{\omega}^m)\,,
% \end{align}
% where $\mathbf{L}\triangleq[\,-\frac{M-1}{2},\dots,\frac{M-1}{2} ]\,^\top$.
% \end{subequations}
%\vspace{-3mm}

\subsection{Derivation of Jacobian Matrix $\mathbf{T}$}
To calculate $\mathbf{T}$, we first define the auxiliary variables $\mathbf{u}_{\mathrm{A}}\triangleq(\mathbf{p}_\text{tx}-\mathbf{p}_\text{ris})/\Vert\mathbf{p}_\text{tx}-\mathbf{p}_\text{ris}\Vert$ and $\mathbf{u}_{\mathrm{D},m}\triangleq(\mathbf{p}_m-\mathbf{p}_\text{ris})/\Vert\mathbf{p}_\text{ms}-\mathbf{p}_\text{ris}\Vert$, 
% \begin{subequations}
% \begin{equation}\label{eq:ua}
%  \mathbf{u}_{\mathrm{A}}= \frac{\mathbf{p}_\text{tx}-\mathbf{p}_\text{ris}}{\Vert\mathbf{p}_\text{tx}-\mathbf{p}_\text{ris}\Vert}, \quad \mathbf{u}_{\mathrm{D},m}= \frac{\mathbf{p}_\text{m}-\mathbf{p}_\text{ris}}{\Vert\mathbf{p}_\text{ms}-\mathbf{p}_\text{ris}\Vert}, 
% \end{equation}
% \begin{equation}\label{eq:ud}
%  \mathbf{u}_{\mathrm{D},m}= \frac{\mathbf{p}_\text{m}-\mathbf{p}_\text{ris}}{\Vert\mathbf{p}_\text{ms}-\mathbf{p}_\text{ris}\Vert}\quad,   
% \end{equation}
% \end{subequations}
as well as $\mathbf{R}_{\alpha} = [\mathbf{r}_1,\mathbf{r}_2,\mathbf{r}_3]^\top$ in \eqref{eq:rot-matrix} with $\mathbf{r}_1 \triangleq [\cos{\alpha},-\sin{\alpha},0]^\top$, $\mathbf{r}_2 \triangleq [\sin{\alpha},\cos{\alpha},0]^\top$, and $\mathbf{r}_3 \triangleq [0,0,1]^\top$. 
Using these variables, we can rewrite AOA and AODs as follows \cite[Appendix A]{nazari2022mmwave}:
\begin{subequations}
\begin{equation}\label{eq:AoA_re}
 \boldsymbol{\phi}= [\,\text{atan2}(\,\mathbf{r}_2^\top\mathbf{u}_{\mathrm{A}},\mathbf{r}_1^\top\mathbf{u}_{\mathrm{A}})\,,\text{acos}(\,\mathbf{r}_3^\top\mathbf{u}_{\mathrm{A}})\,]\,^\top,   
\end{equation}
\begin{equation}\label{eq:AoD_re}
 \boldsymbol{\theta}_m= [\,\text{atan2}(\,\mathbf{r}_2^\top\mathbf{u}_{\mathrm{D},m},\mathbf{r}_1^\top\mathbf{u}_{\mathrm{D},m})\,,\text{acos}(\,\mathbf{r}_3^\top\mathbf{u}_{\mathrm{D},m})\,]\,^\top, 
\end{equation}
\end{subequations}
In the sequel, we compute the derivatives:
\begin{subequations}
\begin{equation}\label{eq:partial_ua_pris}
   \frac{\partial \mathbf{u}_{\mathrm{A}}}{\partial\mathbf{p}_\text{ris} } = (\, \mathbf{u}_{\mathrm{A}}  \mathbf{u}_{\mathrm{A}}^\top-\mathbf{I}_3)\,/\Vert\mathbf{p}_\text{tx}-\mathbf{p}_\text{ris}\Vert,
\end{equation}
\begin{equation}\label{eq:partial_ud_pris}
    \frac{\partial \mathbf{u}_{\mathrm{D},m}}{\partial\mathbf{p}_\text{ris} } = (\, \mathbf{u}_{\mathrm{D},m}  \mathbf{u}_{\mathrm{D},m}^\top-\mathbf{I}_3)\,/\Vert\mathbf{p}_m-\mathbf{p}_\text{ris}\Vert,
\end{equation}
\begin{equation}\label{eq:partial_phiaz_ua}
    \frac{\partial [\boldsymbol{\phi}]_\text{az}}{\partial\mathbf{u}_{A}} = \frac{(\,\mathbf{r}_1^\top\mathbf{u}_{\mathrm{A}})\,\mathbf{r}_2 - (\,\mathbf{r}_2^\top\mathbf{u}_{\mathrm{A}})\,\mathbf{r}_1}{(\,\mathbf{r}_1^\top\mathbf{u}_{\mathrm{A}})\,^2 + (\,\mathbf{r}_2^\top\mathbf{u}_{\mathrm{A}})\,^2},
\end{equation}
\begin{equation}\label{eq:partial_phiel_ua}
    \frac{\partial [\boldsymbol{\phi}]_\text{el}}{\partial\mathbf{u}_{A}} = \frac{-\mathbf{r}_3}{\sqrt{(1-\,\mathbf{r}_3^\top\mathbf{u}_{\mathrm{A}})\,}}, 
\end{equation}
\begin{equation}\label{eq:partial_thetaaz_ua}
    \frac{\partial [\boldsymbol{\theta}_m]_\text{az}}{\partial\mathbf{u}_{\mathrm{D},m}} = \frac{(\,\mathbf{r}_1^\top\mathbf{u}_{\mathrm{D},m})\,\mathbf{r}_2 - (\,\mathbf{r}_2^\top\mathbf{u}_{\mathrm{D},m})\,\mathbf{r}_1}{(\,\mathbf{r}_1^\top\mathbf{u}_{\mathrm{D},m})\,^2 + (\,\mathbf{r}_2^\top\mathbf{u}_{\mathrm{D},m})\,^2},
\end{equation}
\begin{equation}\label{eq:partial_thtael_ua}
    \frac{\partial [\boldsymbol{\theta}_m]_\text{el}}{\partial\mathbf{u}_{\mathrm{D},m}} = \frac{-\mathbf{r}_3}{\sqrt{(1-\,\mathbf{r}_3^\top\mathbf{u}_{\mathrm{D},m})\,}}\quad,
\end{equation}
% \begin{equation}\label{eq:partial_AOA_el_alpha}
%     \frac{\partial [\boldsymbol{\phi}]_{el}}{\partial\alpha} = 0 \qquad ,  \frac{\partial [\boldsymbol{\theta}_m]_{el}}{\partial\alpha} = 0 \quad,
% \end{equation}
\begin{equation}\label{eq:partial_AOA_phi_alpha}
    \frac{\partial [\boldsymbol{\phi}]_\text{az}}{\partial\alpha} = \frac{(\,\mathbf{r}_1^\top\mathbf{u}_{\mathrm{A}})\,{\mathbf{r}^\prime_2}^\top\mathbf{u}_{\mathrm{A}} - (\,\mathbf{r}_1^\top\mathbf{u}_{\mathrm{A}})\,{\mathbf{r}^\prime_1}^\top\mathbf{u}_{\mathrm{A}}}{(\,\mathbf{r}_1^\top\mathbf{u}_{\mathrm{A}})\,^2 + (\,\mathbf{r}_2^\top\mathbf{u}_{\mathrm{A}})\,^2}\quad,
\end{equation}
\begin{equation}\label{eq:partial_AOD_phi_alpha}
    \frac{\partial [\boldsymbol{\theta}_m]_\text{az}}{\partial\alpha} =\frac{(\,\mathbf{r}_1^\top\mathbf{u}_{\mathrm{D},m})\,{\mathbf{r}^\prime_2}^\top\mathbf{u}_{\mathrm{D},m} - (\,\mathbf{r}_1^\top\mathbf{u}_{\mathrm{D},m})\,{\mathbf{r}^\prime_1}^\top\mathbf{u}_{\mathrm{D},m}}{(\,\mathbf{r}_1^\top\mathbf{u}_{\mathrm{D},m})\,^2 + (\,\mathbf{r}_2^\top\mathbf{u}_{\mathrm{D},m})\,^2},
\end{equation}
\end{subequations}
where $\mathbf{r}^\prime_1\triangleq  [-\sin{\alpha},-\cos{\alpha},0]^\top$, $\mathbf{r}^\prime_2\triangleq [\cos{\alpha},-\sin{\alpha},0]^\top$, and $\mathbf{r}^\prime_3\triangleq[0 , 0 , 0]^\top$. Using \eqref{eq:w0} and \eqref{eq:w1}, we can easily find ${\partial \omega_0^m}/{\partial[\boldsymbol{\phi}]_\text{az}}$, ${\partial \omega_0^m}/{\partial[\boldsymbol{\phi}]_\text{el}}$, ${\partial \omega_1^m}/{\partial[\boldsymbol{\phi}]_\text{az}}$, ${\partial \omega_1^m}/{\partial[\boldsymbol{\phi}]_\text{el}}$, ${\partial \omega_0^m}/{\partial[\boldsymbol{\theta}_m]_\text{az}}$, ${\partial \omega_0^m}/{\partial[\boldsymbol{\theta}_m]_\text{el}}$, ${\partial \omega_1^m}/{\partial[\boldsymbol{\theta}_m]_\text{az}}$, and ${\partial \omega_1^m}/{\partial[\boldsymbol{\theta}_m]_\text{el}}$ , which we have not brought in the paper due to page limit.  
% Using \eqref{eq:w0} and \eqref{eq:w1}, yields: 
% \begin{subequations} \label{eq:w_partial}
% \begin{equation}\label{eq:partial_w0_az_alpha}
%  \frac{\partial \omega_0^m}{\partial[\boldsymbol{\phi}]_\text{az}} =  -\sin{\boldsymbol{[\phi}]_\text{el}}\sin{[\boldsymbol{\phi}]_\text{az}} ,  \frac{\partial \omega_0^m}{\partial[\boldsymbol{\phi}]_\text{el}} =  \cos{\boldsymbol{[\phi}]_\text{el}}\cos{[\boldsymbol{\phi}]_\text{az}} , 
% \end{equation}
% \begin{equation}\label{eq:partial_w0_el_alpha}
%  \frac{\partial \omega_0^m}{\partial[\boldsymbol{\theta}_m]_\text{az}} =  -\sin{\boldsymbol{[\theta}_m]_\text{el}}\sin{[\boldsymbol{\theta}_m]_\text{az}} ,
% \end{equation}
% \begin{equation}\label{eq:partial_w1_az_alpha}
%  \frac{\partial \omega_1^m}{\partial[\boldsymbol{\phi}]_\text{az}} =  \sin{\boldsymbol{[\phi}]_\text{el}}\cos{[\boldsymbol{\phi}]_\text{az}},
% % \end{equation}
% % \begin{equation}\label{eq:partial_w1_el_alpha}
%  \frac{\partial \omega_1^m}{\partial[\boldsymbol{\theta}_m]_\text{az}} =  \sin{\boldsymbol{[\theta}_m]_\text{el}}\cos{[\boldsymbol{\theta}_m]_\text{az}}.
% \end{equation}
% \end{subequations}
Using the variables in~\eqref{eq:AoA_re}--\eqref{eq:partial_AOD_phi_alpha}, results in:
\begin{subequations} \label{eq:T_elements}
\begin{equation}\label{eq:partial_tau_pris}
    \mathbf{T}_{m,1:3} =\frac{\partial{\tau}_m}{\partial\boldsymbol{p}_\text{ris}} =  \frac{\mathbf{u}_{\mathrm{A}} + \mathbf{u}_{\mathrm{D},m}}{c} \quad\forall m=\{1, \dots,M\},
\end{equation}
\begin{align}
\label{eq:partial_w0_pris}
    \mathbf{T}_{m,1:3} &=  \frac{\partial{\omega}_0^m}{\partial\mathbf{p}_\text{ris}} =  \left(\,\frac{\partial \omega_0^m}{\partial[\boldsymbol{\phi}]_\text{az}}  \frac{\partial [\boldsymbol{\phi}]_\text{az}}{\partial\mathbf{u}_{A}}+\frac{\partial \omega_0^m}{\partial[\boldsymbol{\phi}]_\text{el}}  \frac{\partial [\boldsymbol{\phi}]_\text{el}}{\partial\mathbf{u}_{A}} \right) \, \frac{\partial \mathbf{u}_{\mathrm{A}}}{\partial\mathbf{p}_\text{ris} } \nonumber \\   & +\left(\frac{\partial \omega_0^m}{\partial[\boldsymbol{\theta}_m]_\text{el}}  \frac{\partial [\boldsymbol{\theta}_m]_\text{el}}{\partial\mathbf{u}_{\mathrm{D},m}} + \frac{\partial \omega_0^m}{\partial[\boldsymbol{\theta}_m]_\text{el}}  \frac{\partial [\boldsymbol{\theta}_m]_\text{el}}{\partial\mathbf{u}_{\mathrm{D},m}}\right)  \frac{\partial \mathbf{u}_{\mathrm{D},m}}{\partial\mathbf{p}_\text{ris} } \nonumber\\
        &\qquad\forall m=\{1+M, \dots,2M\},
\end{align}
\begin{align}
\label{eq:partial_w1_pris}
       \mathbf{T}_{m,1:3} &=  \frac{\partial{\omega}_1^m}{\partial\mathbf{p}_\text{ris}} =  \left(\,\frac{\partial \omega_1^m}{\partial[\boldsymbol{\phi}]_\text{az}}  \frac{\partial [\boldsymbol{\phi}]_\text{az}}{\partial\mathbf{u}_{A}}+\frac{\partial \omega_1^m}{\partial[\boldsymbol{\phi}]_\text{el}}  \frac{\partial [\boldsymbol{\phi}]_\text{el}}{\partial\mathbf{u}_{A}} \right) \, \frac{\partial \mathbf{u}_{\mathrm{A}}}{\partial\mathbf{p}_\text{ris} } \nonumber \\   & +\left(\frac{\partial \omega_1^m}{\partial[\boldsymbol{\theta}_m]_\text{el}}  \frac{\partial [\boldsymbol{\theta}_m]_\text{el}}{\partial\mathbf{u}_{\mathrm{D},m}} + \frac{\partial \omega_1^m}{\partial[\boldsymbol{\theta}_m]_\text{el}}  \frac{\partial [\boldsymbol{\theta}_m]_\text{el}}{\partial\mathbf{u}_{\mathrm{D},m}}\right)  \frac{\partial \mathbf{u}_{\mathrm{D},m}}{\partial\mathbf{p}_\text{ris} } \nonumber\\
        &\qquad\forall m=\{1+2M, \dots,3M\},
\end{align}
\end{subequations}
Finally, considering ~\eqref{eq:partial_AOA_phi_alpha}--\eqref{eq:partial_AOD_phi_alpha}, we have:
\begin{subequations}
\begin{align}
\label{eq:partial_w0_alpha}
    \mathbf{T}_{m,4} =  \frac{\partial{\omega}_0^m}{\partial\alpha} =  \frac{\partial \omega_0^m}{\partial[\boldsymbol{\phi}]_\text{az}}  \frac{\partial [\boldsymbol{\phi}]_\text{az}}{\partial\alpha}+ \frac{\partial \omega_0^m}{\partial[\boldsymbol{\theta}_m]_\text{az}}  \frac{\partial [\boldsymbol{\theta}_m]_\text{az}}{\partial\alpha} \nonumber \\  
        \quad\forall m=\{1+M, \dots,2M\},
\end{align}
\begin{align}
\label{eq:partial_w1_alpha}
     \mathbf{T}_{m,4}=  \frac{\partial{\omega}_1^m}{\partial\alpha} =  \frac{\partial \omega_1^m}{\partial[\boldsymbol{\phi}]_\text{az}}  \frac{\partial [\boldsymbol{\phi}]_\text{az}}{\partial\alpha}+ \frac{\partial \omega_1^m}{\partial[\boldsymbol{\theta}_m]_\text{az}}  \frac{\partial [\boldsymbol{\theta}_m]_\text{az}}{\partial\alpha} \nonumber\\  
        \quad\forall m=\{1+2M, \dots,3M\},
\end{align}
\end{subequations}
and the remaining elements of $\mathbf{T}$ are zero.
\balance 
\bibliographystyle{IEEEtran}
\bibliography{ref}
\end{document}